\documentclass{Rinton-P9x6}
\usepackage{amsmath,amsfonts,amscd,amsbsy}
\title{Operation Time of Quantum Gates}
\date{}
\author{Lev B.\ Levitin, Tommaso Toffoli, Zachary Walton}
\address{Boston University. ECE Dept., 8 Mary's St., Boston MA 02215\\
E-mails: {\tt levitin@bu.edu}, {\tt tt@bu.edu}, {\tt walton@bu.edu}}

 \makeatletter

 \DeclareRobustCommand\em
        {\@nomath\em \ifdim \fontdimen\@ne\font >\z@
                       \upshape \else \slshape \fi}

\def\@begintheorem#1#2{\sl \trivlist \item[\hskip \labelsep{\bf #1\ #2}]}
\def\@opargbegintheorem#1#2#3{\sl \trivlist
     \item[\hskip \labelsep{\bf #1\ #2\ (#3)}]}

 \mathchardef\BY="0202

 \def\@empty{}
 \newcommand{\asin}[2][]{{
    \def\t@mp{#1}%
    \def\@cite##1##2{\marginpar{\hfil{\footnotesize$
    \ifx\t@mp\@empty\text{##2}\else\frac{\text{##1}}{\text{##2}}\fi$}\hfil}}%
\cite[#1]{#2}}}

 \newcommand{\sectlabel}[1]{\label{sect:#1}}
 
 \newcommand{\eqlabel}[1]{\label{eq:#1}}

 \newcommand{\Chapt}[2][]{\def\t@mp{#1}%
\chapter{#2} \ifx\t@mp\@empty\else\sectlabel{#1}\fi}
 \newcommand{\Sect}[2][]{\def\t@mp{#1}%
\section{#2} \ifx\t@mp\@empty\else\sectlabel{#1}\fi}
 \newcommand{\Subsect}[2][]{\def\t@mp{#1}%
\subsection{#2} \ifx\t@mp\@empty\else\sectlabel{#1}\fi}
 
 \newcommand{\Eq}[2][]{\def\t@mp{#1}%
\begin{equation}#2\ifx\t@mp\@empty\notag\else\eqlabel{#1}\fi\end{equation}}
 \newcommand{\Eqaligned}[2][]{\def\t@mp{#1}%
\begin{equation}\begin{aligned}#2\end{aligned}
\ifx\t@mp\@empty\notag\else\eqlabel{#1}\fi
\end{equation}}

 \newcommand{\sect}[1]{\S\ref{sect:#1}}      

 \newcommand{\eq}[1]{(\ref{eq:#1})}	

 \long\def\endsubsection#1{\smallskip\hbox to\hsize{\leaders\hrule\hfill\ \sect{#1}}\medskip}

  \def\@arabic#1{\number #1} 

\long\def\@makecaption#1#2{
	\vskip\abovecaptionskip
	\sbox\@tempboxa{{\small #1: #2}}%
	\ifdim\wd\@tempboxa>\hsize
	    {\small #1: #2\par}
	\else
	   \global\@minipagefalse
	   \hbox to\hsize{\hfil\box\@tempboxa\hfil}
	\fi
	\vskip\belowcaptionskip}

\def\figstrut#1{\hbox to\linewidth{\vrule height#1\hfill}}

\def\cstrip#1{\setbox0=\hbox{$#1$}\kern-.5\wd0\lower2pt\box0}
\def\rstrip#1{\setbox0=\hbox{$#1$}\kern-\wd0\lower2pt\box0}
\def\lstrip#1{\setbox0=\hbox{$#1$}\lower2pt\box0}
\def\tstrip#1{\setbox0=\hbox{$#1$}\kern-.5\wd0\lower\ht0\box0}
\def\bstrip#1{\setbox0=\hbox{$#1$}\kern-.5\wd0\raise\ht0\box0}
\def\Lstrip#1{\setbox0=\hbox{$\mskip2mu#1$}\lower2pt\box0}

\def\idpad{\thinspace}
\def\id{\idpad\begingroup \tt \let\do\@makeother \dospecials 
          \@ifstar{\@sid}{\@id}}
\def\@sid#1{\def\@tempa ##1#1{##1\endgroup\idpad}\@tempa}
\def\@id{\obeyspaces \frenchspacing \@sid}

\makeatother

 \newenvironment{tbmatrix}{\bigl[\begin{smallmatrix}}{\end{smallmatrix}\bigr]}
 \newcommand{\twomatrix}[4]{\begin{bmatrix}#1&#2\\#3&#4\end{bmatrix}}

 \newcommand{\ttwovector}[2]{\begin{tbmatrix}#1\\#2\end{tbmatrix}}
 \newcommand{\threematrix}[9]{\begin{bmatrix}#1&#2&#3\\#4&#5&#6\\#7&#8&#9\end{bma
trix}}

\def\H{\mathbf{H}}
\def\I{\mathbf{I}}
\def\U{\mathbf{U}}
\def\pauli{\boldsymbol{\sigma}}

\begin{document}

\maketitle

\abstracts{We consider a quantum gate that complements the state of a qubit
and then adds to it an arbitrary phase shift. It is shown that the minimum
operation time of the gate is $\tau = \frac h{4E}(1+2\frac\theta\pi)$, where
$h$ is Planck's constant, $E$ is the quantum-mechanical average energy, and
$\theta$ is the phase shift modulo $\pi$.}

\noindent It had been shown in \cite{margolus/levitin} that there exists a
fundamental limit to the speed of dynamical evolution of a quantum
system. Namely, the minimum time required for a system to go from a given
state to one orthogonal to it is
 \Eq[time-orth]{\tau=\frac h{4E},}
 where $h$ is Planck's constant and $E$ is the quantum-mechanical average
energy of the system. Expression \eq{time-orth} applies to the
\emph{autonomous} time evolution of a system, and it is not immediately
applicable to changes in the system state caused by the interaction with
another (external) system.

This paper considers the question of what is the minimum time of operation of
quantum gates that operate on qubits (i.e., quantum systems with
two-dimensional Hilbert space).

\medskip

Let
 \Eq[two_states]{\psi_1(0) = \ttwovector01\quad\text{and}\quad\psi_2(0) = \ttwovector10}
 be the two initial orthogonal stationary states of a qubit.  Consider a
``gate'' that complements the state of the qubit (a quantum inverter or a
controlled-{\sc not} gate with the controlling qubit in logical state 1) and
then adds to it an arbitrary phase shift $\theta$. This is a device that
applies an external interaction to the system for a certain time $\tau$ such
that at the end of this time interval
 \Eq[invert]{\psi_1(\tau)=\psi_2(0)e^{-i\theta}\quad\text{and}\quad \psi_2(\tau)=\psi_1(0)e^{-i\theta}}
 i.e., the two orthogonal states are swapped and a given phase shift
$\theta$ is added to the resulting state.
 During this time the evolution of the system is governed by a unitary
operator $\U(t)$, so that
 \Eq[unitary]{\psi_i(t) = \U(t)\psi_i(0) = e^{-\frac{i\H}{\hbar} t}\psi_i(0),\quad (i=1,2),}
 where $\H$ is the total Hamiltonian of the system (including the interaction
Hamiltonian) and $\hbar=h/2\pi$. Note that, owing to linearity, \eq{invert}
is a necessary and sufficient condition for an \emph{unknown} quantum state
$a\psi_1(0)+b\psi_2(0)$ of a qubit to be converted into the ortogonal state
with a phase shift $\theta$, provided that $\text{Re}(ab^*)=0$ (this
condition specifies a two-parameter family of states). Note also that the
overall phase of the state is essential, since this qubit is intended
to be part of a many-qubit system.

Recall that any unitary time evolution operator has a diagonal form,
 \Eq[diagonal]{\U_\text{diag}(t)
	 = \twomatrix{e^{-i\frac{E_1}\hbar t}}00{e^{-i\frac{E_2}\hbar t}},}
 where $E_1$, $E_2$ are the eigenvalues of $\H$. The most general form
of a unitary operator with the diagonal form \eq{diagonal}
can be written as follows:
 \Eqaligned{
    &\U(t) = e^{-i\frac{E_1+E_2}{2\hbar}t}\times\\
    &\ \times\twomatrix
{\cos^2\!\phi_1\,e^{i\frac{E_2-E_1}{2\hbar}t}+\sin^2\!\phi_1\,e^{-i\frac{E_2-E_1}{2\hbar} t}}
{2ie^{i\phi_2}\sin\phi_1\cos\phi_1 \sin\frac{E_2-E_1}{2\hbar}t}
{2ie^{-i\phi_2}\sin\phi_1\cos\phi_1 \sin\frac{E_2-E_1}{2\hbar}t}
{\cos^2\!\phi_1\,e^{-i\frac{E_2-E_1}{2\hbar}t}+\sin^2\!\phi_1\,e^{i\frac{E_2-E_1}{2\hbar} t}},
 }
 where $\phi_1$ and $\phi_2$ are arbitrary phase parameters. (Note that, in
fact, $\U(t)$ has three independent parameters---not four.)

It is readily shown that, to satisfy \eq{invert} and to provide for the
minimum time $\tau$, one should choose $\phi_1=\frac\pi4$ and $\phi_2=0$.
Hence the operator $\U(t)$ has---in the original basis of the system (the one
with basis vectors $\psi_1(0)$ and $\psi_2(0)$)---the following unique form
 \def\angle{\frac{E_2-E_1}{2\hbar}t}
 \Eq[main_unitary]{\U(t)= e^{-i\frac{E_1+E_2}{2\hbar}t}
	\twomatrix{\cos\angle\quad}{i\sin\angle}{i\sin\angle\quad}{\cos\angle}.}
 The corresponding Hamiltonian is
 \Eq[main_hamilton]{\H=
\twomatrix{\frac{E_1+E_2}2\ }{\frac{E_1-E_2}2}{\frac{E_1-E_2}2\ }{\frac{E_1+E_2}2}
= \frac{E_1+E_2}2\I + \frac{E_1-E_2}2\pauli_x,}
 where $\I$ is the identity operator and $\pauli_x$ one of the Pauli
matrices. It follows from \eq{unitary} and \eq{main_unitary} that, for a proper
choice of $E_1$ and $E_2$,
 \Eqaligned[swap]{\psi_1(\tau) = \U(\tau)\psi_1(0) = \psi_2(0)e^{-i\theta},\\
		  \psi_2(\tau) = \U(\tau)\psi_2(0) = \psi_1(0)e^{-i\theta}.}
 Equation \eq{swap} holds if $E_1$ and $E_2$ are chosen so that
 \Eqaligned[Epair]{
	E_1 &= \frac\theta{\theta+\pi}E_2 &\text{for}\quad &0\leq\theta<\pi,\quad\text{and}\\
	E_2 &= \frac{\theta-\pi}\theta E_1 &\text{for}\quad &\pi\leq\theta<2\pi.
 }
 Then the minimum time $\tau$ of operation of the quantum gate is given by
 \Eq[min_time]{\tau = \frac h{2|E_2-E_1|}.}
 The average energy of the system is
 \Eq[averageE]{E = \langle\psi_i|\H|\psi_i\rangle = \frac{E_1+E_2}2,\quad(i=1,2).}
 From \eq{Epair}, \eq{min_time}, and \eq{averageE} it follows that
 \Eq[delay]{
   \tau=\tau(\theta)=\frac h{4E}\left(1+2\cdot\frac{\theta\bmod\pi}\pi\right).}

It may seem that $\tau$ could be made arbitrarily small by increasing the
energy difference $|E_2-E_1|$. However, this is not possible in a real
physical situation, which is not completely described by the qubit
model. For example, if $\psi_1(0)$ and $\psi_2(0)$ are two
stationary states of an electron in an atom or ion, with energies
$E_1^\prime$ and $E_2^\prime$ respectively, then it is advantageous to let
$E_2-E_1$ be equal to the resonant frequency,
 \Eq{E_2-E_1 = E_2^\prime - E_1^\prime}
 (cf.\ \cite[Ch.\ 6]{ll}). As a numerical example, consider experiments
\cite{steane} made with Ca$^+$ ions in an ion trap. The characteristic
wavelength of the transition between the two relevant Ca$^+$ energy levels is
$\lambda = 397\,\text{nm}$, which yields $\tau = \tfrac \lambda{2c} \sim
6.62\cdot10^{-16}\,\text{s}$.

\medskip

 Consider now a one-qubit quantum gate that makes an arbitrary unitary
transformation of a \emph{known} state such that the absolute value of the
inner product
 \Eq{\big|\langle\psi(\tau_\alpha)|\psi(0)\rangle\big| = \cos\alpha.}
 A similar analysis shows that, for any $\alpha$ ($0\leq\alpha\leq\tfrac\pi2$),
the minimum time required for this operation is
 \Eq[final]{\tau_\alpha = \frac{\alpha h}{2\pi E} = \frac{2\alpha}\pi
\tau(0).}

\end{document}